\documentclass[12pt]{article}

\RequirePackage{etex}

\usepackage[margin=1in]{geometry}  	
\usepackage{amsmath,amsfonts,amssymb,amsthm}
\usepackage{booktabs}
\usepackage{multirow}
\usepackage[onehalfspacing]{setspace} 
\renewcommand{\arraystretch}{0.6} 

\usepackage{amsmath}
\usepackage{tcolorbox}

\usepackage{float}
\usepackage{caption}
\usepackage{graphicx}
\usepackage{adjustbox}
\usepackage{lscape}

\usepackage{adjustbox}

\usepackage{xcolor}

\usepackage{relsize}

\usepackage{bbm}
\usepackage{subfig}

\usepackage[sc]{titlesec}

\usepackage{authblk}

\usepackage{amsthm}

\newtheorem*{proposition*}{Proposition}

\usepackage{xpatch}
\makeatletter
\xpatchcmd{\proof}{\@addpunct{.}}{\@addpunct{:}}{}{}
\makeatother

\makeatletter
\newcommand{\vast}{\bBigg@{3}}
\newcommand{\Vast}{\bBigg@{4}}
\makeatother

\newcommand\independent{\protect\mathpalette{\protect\independenT}{\perp}}
\def\independenT#1#2{\mathrel{\rlap{$#1#2$}\mkern2mu{#1#2}}}

\makeatletter
\newcommand*{\indep}{%
  \mathbin{%
    \mathpalette{\@indep}{}%
  }%
}
\newcommand*{\nindep}{%
  \mathbin{
    \mathpalette{\@indep}{\not}
  }%
}
\newcommand*{\@indep}[2]{%
  \sbox0{$#1\perp\m@th$}
  \sbox2{$#1=$}
  \sbox4{$#1\vcenter{}$}
  \rlap{\copy0}
  \dimen@=\dimexpr\ht2-\ht4-.2pt\relax
  \kern\dimen@
  {#2}%
  \kern\dimen@
  \copy0 
} 
\makeatother

\DeclareMathOperator{\E}{\textnormal{\mbox{E}}}

\usepackage[utf8]{inputenc}
\DeclareUnicodeCharacter{200E}{}

\usepackage{setspace}
\usepackage{cite}

\usepackage{xr-hyper}

\usepackage{xcolor}
\definecolor{forestgreen}{RGB}{34,139,34}
\usepackage{hyperref}
\hypersetup{
    colorlinks=true,
    linkcolor=magenta,
    filecolor=magenta, 
    citecolor=forestgreen,      
    urlcolor=blue
}

\makeatletter
\renewcommand{\boxed}[1]{\text{\fboxsep=.2em\fbox{\m@th$\displaystyle#1$}}}
\makeatother

\makeatletter
\newcommand*{\addFileDependency}[1]{
  \typeout{(#1)}
  \@addtofilelist{#1}
  \IfFileExists{#1}{}{\typeout{No file #1.}}
}
\makeatother

\newcommand*{\myexternaldocument}[1]{%
    \externaldocument{#1}%
    \addFileDependency{#1.tex}%
    \addFileDependency{#1.aux}%
}

\myexternaldocument{appendix}

\usepackage[stable]{footmisc}

\usepackage{rotating}

\makeatletter
\def\@hangfrom#1{\setbox\@tempboxa\hbox{{#1}}%
      \hangindent 0pt
      \noindent\box\@tempboxa}
\makeatother

\makeatletter
\def\@seccntformat#1{\@ifundefined{#1@cntformat}%
   {\csname the#1\endcsname\quad}  
   {\csname #1@cntformat\endcsname}
}
\let\oldappendix\appendix 
\renewcommand\appendix{%
    \oldappendix
    \newcommand{\section@cntformat}{\appendixname~\thesection\quad}
}
\makeatother

\usepackage[absolute,showboxes]{textpos}

\TPMargin{5pt}

\newcommand{\copyrightstatement}{
    \begin{textblock}{0.84}(0.08,0.93)    
         \noindent
         \footnotesize
         This DRAFT manuscript presents WORK IN PROGRESS. Please send comments to \href{mailto:idahabreh@hsph.harvard.edu}{idahabreh@hsph.harvard.edu}.
    \end{textblock}
}

\usepackage{datetime}
\def\paperversionmajor{9}
\def\paperversionminor{0}

\begin{document}

\copyrightstatement

\title{Selection on treatment in the target population of generalizabillity and transportability analyses \vspace*{0.3in} }

\author[1,2]{Yu-Han Chiu}
\author[1-3]{Issa J. Dahabreh
}

\affil[1]{CAUSALab, Harvard T.H. Chan School of Public Health, Boston, MA}
\affil[2]{Department of Epidemiology, Harvard T.H. Chan School of Public Health, Boston, MA}
\affil[3]{Department of Biostatistics, Harvard T.H. Chan School of Public Health, Boston, MA}

\maketitle{}
\thispagestyle{empty}

\newpage
\thispagestyle{empty}

\vspace*{1in}
{\LARGE \centering Selection on treatment in the target population of generalizability and transportability analyses \par }

\vspace{1in}
\noindent
\textbf{Running head:} Selection on treatment in generalizability and transportability analyses

\vspace{0.3in}
\noindent
\textbf{Word count:} abstract = 198; main text = ~4000.

\vspace{0.3in}
\noindent
\textbf{Abbreviations that appear in the text:} None.

\clearpage
\thispagestyle{empty}

\vspace*{1in}

\begin{abstract}
\noindent
\linespread{1.4}\selectfont
Investigators are increasingly using novel methods for extending (generalizing or transporting) causal inferences from a trial to a target population. In many generalizability and transportability analyses, the trial and the observational data from the target population are separately sampled, following a non-nested trial design. In practical implementations of this design, non-randomized individuals from the target population are often identified by conditioning on the use of a particular treatment, while individuals who used other candidate treatments for the same indication or individuals who did not use any treatment are excluded. In this paper, we argue that conditioning on treatment in the target population changes the estimand of generalizability and transportability analyses and potentially introduces serious bias in the estimation of causal estimands in the target population or the subset of the target population using a specific treatment. Furthermore, we argue that the naive application of marginalization-based or weighting-based standardization methods does not produce estimates of any reasonable causal estimand. We use causal graphs and counterfactual arguments to characterize the identification problems induced by conditioning on treatment in the target population and illustrate the problems using simulated data. We conclude by considering the implications of our findings for applied work. \\

\vspace{0.2in}
\noindent
\textbf{Keywords:} randomized trials; observational analyses; transportability; generalizability; data linkage; combining information.

\end{abstract}

\clearpage
\section*{Introduction}
\setcounter{page}{1}

\noindent
Estimates of counterfactual probabilities and treatment effects from a trial may not directly apply to the population where the trial results will be used to inform decisions when that population has a  different distribution of variables that predict the outcome or modify the effect of treatment compared to the trial. To address this issue, investigators are increasingly using methods to extend -- generalize or transport \cite{dahabreh2019commentaryonweiss} -- causal inferences from trials to new populations of substantive interest (e.g., \cite{cole2010, westreich2017, rudolph2017, dahabreh2018generalizing, dahabreh2020transportingStatMed}). These methods use data on baseline covariates, treatments, and outcomes from the trial, but only use data on baseline covariates from the new population; thus, the methods do not require the assumption that there is no confounding outside the trial. 

For many generalizability and transportability analyses, trial data and data on non-randomized individuals are separately sampled using a non-nested trial design \cite{dahabreh2021studydesigns}. For instance, many analyses combine data from a completed trial with a separately obtained sample of non-randomized individuals identified in routinely collected data (e.g., health insurance claims or electronic health records). In such analyses, investigators are interested in learning about the population underlying the the sample of non-randomized individuals; hereafter, we refer to this population as the \emph{target population}. Non-nested trial designs face the practical challenge of how to sample the target population. Many investigators attempt to address this challenge by (1) identifying individuals in the target population who used some treatment(s) of interest (usually one or both of the treatments examined in the trial) with the same indication as the treatments in the trial; and (2) excluding individuals who did not use any treatment or those who used other treatments for the same indication. 

Restricting the target population sample to individuals who use some treatment(s) of interest appears to naturally follow from standard strategies for emulating a target trial by comparing initiators of the treatments of interest, and excluding individuals who initiate other treatments or who do not initiate treatment. In this paper, however, we argue that conditioning on treatment in the target population changes the estimand of interest and potentially introduces serious bias in studies extending inferences from a trial to a target population. Furthermore, we argue that the statistical quantities estimated by applying popular marginalization-based or weighting-based standardization methods \cite{dahabreh2019relation} after conditioning on treatment may not correspond to any reasonable causal estimand. To our knowledge, these issues have not received attention in the applied literature on generalizability and transportability analyses; in fact, conditioning on treatment in the target population appears to be common, both in applied and methodological work (e.g.,  \cite{berkowitz2018generalizing, hong2019comparison, webster2019diagnostic, webster_clark2020, lund2020effectiveness, mollan2021transportability, wang2022approaches}). We characterize the identification problems induced by conditioning on treatment using causal graphs and counterfactuals and we illustrate how they arise using simulated data. Last, we consider the implications of our findings for future analyses extending inferences from a trial to a target population.

\section*{Example: Anticoagulants for Atrial Fibrillation} 

To fix ideas, we consider the example of treatment with novel oral anticoagulants for patients with atrial fibrillation. In the United States, dabigatran was the first novel oral anticoagulant approved by the U.S. Food and Drug Administration in late 2010. Approval was based on the RE-LY randomized trial \cite{chen2020direct} that compared the effect of dabigatran versus warfarin in patients with atrial fibrillation at high-risk of stroke \cite{connolly2009dabigatran}. At 2 years of follow up, the trial found that dabigatran was superior to warfarin for the composite outcome of stroke or systematic embolism. Soon after the approval of dabigatran, individuals with atrial fibrillation could use dabigatran or warfarin. Following the approval of dabigatran, several other novel oral anticoagulants (e.g., rivaroxaban, apixaban, edoxaban, betrixaban) were approved and have been adopted in current clinical practice \cite{chen2020direct}; nowadays there would be multiple possible treatments for individuals with atrial fibrillation and some patients may not receive any treatment. For the moment we will consider only treatment with dabigatran or warfarin in a population where every treatment candidate receives one of these two alternatives. In eAppendix A, we extend our results to consider multivalued treatments (the possibility that some individuals use no treatment is a special case of the results presented in the eAppendix).  

Suppose that a research team wanted to examine the effectiveness of dabigatran compared with warfarin in a ``real-world'' target population of trial-eligible patients. Because trial participants may not have the same distribution of outcome predictors or effect modifiers as non-randomized patients, estimates of counterfactual risk and treatment effects from the trial may not apply directly to the target population. Conducting another trial in a random sample of the target population can be costly, infeasible, and potentially unethical given existing evidence; instead, the investigators could conduct an observational analysis to emulate a target trial similar to the RE-LY trial using routinely collected (e.g., health insurance claims or electronic health records) or registry data from the target population \cite{hernan2016, dahabreh2020benchmarking}. There is some risk, however, that estimates from such an observational analysis would be subject to unmeasured confounding, particularly during the early adoption period for a new medication \cite{schneeweiss2011assessing}. 

To avoid assumptions about the absence of unmeasured confounding in the observational analysis, another approach for assessing the comparative effectiveness of dabigatran versus warfarin in the target population is to extend causal inferences from the trial to the target population. Suppose that the investigators were able to obtain access to the RE-LY trial data and, to sample the target population, they decided to identify trial-eligible individuals in an insurance claims database during a relevant study period (e.g., within 12 months of dabigatran becoming available). Because an individual may meet eligibility criteria multiple times in routinely collected or registry data, there are multiple ways to sample the target population. One popular approach is to select individuals based on the use of some specific treatment(s). For instance, in our example, the investigators might select individuals who initiated warfarin during the study period, as is a new-/incident-user design \cite{ray2003, johnson2013}, and identify their baseline covariates using a covariate assessment period prior to the initiation of treatment. The covariate distribution of this sample of warfarin-initiators could then be used to standardize the trial results using marginalization-based or weighting methods.

In order to examine the problems with this approach for sampling the target population, we introduce some notation and describe a simple causal model for extending inferences from a randomized trial to the target population.

\section*{Causal Models and Estimands}

\paragraph{Notation and basic setup:} Let $A$ denote treatment strategy (0 for warfarin; 1 for dabigatran), $Y$ a binary outcome (i.e., a composite of stroke, systemic embolism, and death measured at the end of study), $S$ the indicator of trial participation (1 yes, 0 no), $X$ the measured baseline covariates, and $U$ unmeasured covariates. In what follows, expectations (and probabilities) are with respect to the sampling model of the non-nested trial design, where trial participants and non-participants are separately sampled into the study \cite{dahabreh2020transportingStatMed, dahabreh2021studydesigns}.

To simplify exposition, we assume perfect adherence to the assigned (recommended) treatment strategy, both in the trial and the observational study, and no loss-to-follow-up. Though unrealistic, these assumptions will allow us to focus on the problems engendered by selection on treatment in the target population.

\paragraph{Causal models:} Throughout, we adopt a non-parametric structural equation model with finest fully randomized causally interpretable structured tree graph errors \cite{robins1986} and represent these models using causal directed acyclic graphs (DAGs) \cite{pearl2009causality, spirtes2000causation}. Under our structural model, counterfactuals \cite{rubin1974, robins2000d} are assumed well-defined for interventions on any variables and consistency holds for these interventions \cite{richardson2013single}. Specifically, the following counterfactual variables are well-defined: $A^{s=1}$, the counterfactual assignment under intervention to scale-up trial engagement by setting $S$ to $s=1$; $Y^{s=1,a}$, the counterfactual outcome under joint intervention to (1) scale up outcome-relevant aspects of trial engagement ($s=1$), and (2) set treatment $A$ to $a$ \cite{dahabreh2019identification}; and $Y^{a}$, the counterfactual outcome under intervention to set treatment to $a$. 

The causal DAG in Panel (A) of Figure 1 depicts the simplified structure for our running example. The measured prognostic factors $X$ may affect trial participation and the outcome, as represented by the $S \leftarrow X \rightarrow Y$ fork. The lack of unmeasured common causes of trial participation $S$ and the outcome $Y$ indicates the typical assumption in generalizability and transportability analyses that trial participants and non-participants are exchangeable with respect to counterfactual outcomes for interventions on treatment $A$, conditional on the covariates $X$. The lack of an $S \rightarrow Y$ edge  encodes the assumption that trial participation affects the outcome only through treatment (i.e., there are no ``trial engagement effects'' \cite{dahabreh2019identification}). In fact, in all the causal models we consider in this paper there is no direct effect of trial participation on the outcome and all effects of trial participation intersect (are mediated by) treatment $A$; thus, we have that $Y^{s=1,a} = Y^{a}$ \cite{dahabreh2019identification}. The $S \rightarrow A$ edge represents that treatment assignment is different between trial participants (randomized; cannot depend on unmeasured variables) and non-participants (can depend on unmeasured covariates). The $A \leftarrow U \rightarrow Y$ fork denotes the presence of confounding by unmeasured variables of treatment assignment (among the non-randomized individuals with $S = 0$).  

Panels (B) and (C) of Figure 1 are causal DAGs by trial participation status; we denote conditioning on $S=1$ in Panel (B) and on $S = 0$ in Panel (C) by placing a rectangle around the $S$ node. Among individuals participating in the trial ($S = 1$), we assume that treatment assignment is marginally randomized (i.e., does not depend on the covariates $X$), reflected in the absence of an $X \rightarrow A$ edge in Panel (B) of Figure 1. Among individuals not participating in the trial ($S = 0$), both measured prognostic factors $X$ and unmeasured prognostic factors $U$ may affect the treatment assignment, represented by the $A \leftarrow X \rightarrow Y$ and $A \leftarrow U \rightarrow Y$ forks, respectively, in Panel (C) of Figure 1.

\paragraph{Causal Estimands:} In non-nested trial designs, a key causal estimand is the expectation of the counterfactual outcome under treatment $a$ in the target population, defined as 
\begin{equation}\label{eq:betadef}
    \beta(a) \equiv \E[Y^{a} | S=0].
\end{equation}

The expectation of the difference of the counterfactual outcomes under the two different treatments, $\E[Y^{a=1} - Y^{a=0} | S=0] = \E[Y^{a=1}|S=0] - \E[Y^{a=0} | S=0] = \beta(1) - \beta(0)$, is the average treatment effect in the target population. It is important to note that these counterfactual expectations and the average treatments effect pertain to individuals in the target population, regardless of the treatment they might actually use. For instance, in our running example, they pertain to the entire target population, regardless of dabigatran or warfarin use.

We may also be interested in causal estimands that pertain to the \emph{subset} of the target population who use some specific treatment, say $a^\prime$. The expectation of the counterfactual outcome in the subset of the target population using treatment $a^\prime$ is defined as 
\begin{equation}\label{eq:gammadef}
    \gamma(a, a') \equiv \E[Y^{a} | S=0, A=a'].
\end{equation}

In our running example, setting $a=1$ and $a'=0$, $\gamma(1, 0)$ is the expectation of the counterfactual outcome under dabigatran treatment ($a=1$) among the subset of the target population using warfarin ($A = a'= 0$). This quantity is the expectation of the counterfactual outcome under the experimental treatment (dabigatran) among individuals in the target population using the more established treatment (warfarin) and thus may be of particular scientific interest. Similarly, using $a=0$ and $a'=1$, $\gamma(0, 1)$ is the expectation of the counterfactual outcome under warfarin (the control treatment) among the dabigatran treated subset of the target population. 

The expectation of the difference of the counterfactual outcomes under the two different treatments in the subset of the target population using treatment $A = a^\prime$, is $\E[Y^{a=1} - Y^{a=0} | S=0, A = a^\prime] = \E[Y^{a=1}|S=0, A = a^\prime] - \E[Y^{a=0} | S=0, A = a^\prime] = \gamma(1, a^\prime) - \gamma(0, a^\prime)$ is a kind of target population average treatment effect ``on the treated'' \cite{hartman2015sample}.

\section*{Identifiability Conditions}

Here we review key identifiability conditions that are true in our set up, and are often invoked in generalizability and transportability analyses.   

\paragraph{Exchangeability conditions:} Under our assumed causal structure, the counterfactual outcome under treatment $a$, $Y^a$, is independent of trial participation, conditional on the baseline covariates: $Y^a \indep S | X$. This condition implies that $\E[Y^a = 1 | X, S = 1] = \E[Y^a = 1 | X, S = 0]$. Furthermore, among trial participants, the counterfactual outcome mean under treatment $a$ is independent of treatment, conditional on baseline covariates: $Y^a \indep A | (X, S = 1)$. This condition implies that $\E[Y^a | X , S = 1] = \E[Y^a | X , S = 1, A = a]$. The two exchangeability conditions can be read off from a single-world intervention graph \cite{richardson2013single} for a joint intervention to scale-up outcome-relevant trial procedures by intervening to set $S$ to $s=1$ and to set $A$ to $a$, provided that the graph incorporates the context-specific knowledge \cite{robins2022interventionist, shpitser2022multivariate} that treatment assignment is independent of both measured and unmeasured covariates for individuals in the trial \cite{dahabreh2019identification, dahabreh2020benchmarking}. For completeness, we construct this graph in Appendix Figure 1.

\paragraph{Positivity conditions:} We assume that $\Pr[S=1|X=x]>0$ for every $x$ with joint density $f(X = x, S = 0) \neq 0$. Furthermore, for each $a \in \{0,1\}$ and each $s \in \{0,1\}$, we assume that $\Pr[A=a | X=x, S =s] > 0$ for every $x$ with joint density $f(X = x, S = s) \neq 0$. 

\paragraph{Consistency conditions:} For every treatment $a \in \{ 0, 1 \}$, if an individual uses treatment $A = a$, their observed outcome equals their counterfactual outcome under hypothetical intervention to set treatment to $a$. That is, if $A = a$, then $Y = Y^a$.

\section*{The Parameter of Standardization Procedures After Conditioning on Treatment}

In a number of studies sampling the target population by conditioning on treatment, and in our running example, investigators are standardizing the trial data to the covariate distribution of a sample from the target population selected on treatment. The target statistical parameter of this common approach can be defined as follows:
\begin{equation}\label{eq:statistical_parameter_general}
    \begin{split}
   \phi(a,a^\prime) &\equiv \E[ \E[Y | X, S= 1, A = a] \big| S = 0, A = a^\prime ] \\
            &= \int \E[Y | X = x, S= 1, A = a] f(x | S = 0, A = a^\prime )dx.
   \end{split}
\end{equation}

In other words, $\phi(a,a^\prime)$ is the large sample limit of applying popular marginalization-based estimators \cite{dahabreh2019relation} to a composite dataset that is constructed by appending the trial data to the sample from the target population  (the latter restricted to individuals who used treatment $A = a^\prime$). This parameter, which is defined in terms of measured variables, involves standardizing the conditional expectation of the observed outcome among individuals assigned to treatment $A = a$ in the trial,  $\E[Y | X = x, S= 1, A = a]$, to the covariate distribution of the subset of the target population who used treatment $A = a'$, $f(x | S = 0, A = a^\prime )$. In our running example, using $a=1$ and $a' = 0$, $\phi(1,0)$ denotes standardization of the dabigatran arm ($a = 1$) of the trial to the warfarin-treated subset of the target population (for whom $A = a' = 0$). 

Though the use of standardization methods is very natural, to the point that it has come to define applied generalizability and transportability analyses, it is not immediately clear that the statistical parameter $\phi(a,a^\prime)$ defined in equation \eqref{eq:statistical_parameter_general} equals either (1) $\beta(a)$, the average treatment effect in the target population in equation \eqref{eq:betadef}, or (2) $\gamma(a, a^\prime)$, the counterfactual outcome mean under treatment $a$  in the subset of the target population who used treatment $a'$ in equation \eqref{eq:gammadef}. In fact, we will next argue that under the causal structure of Figure 1,  $\phi(a,a^\prime)$ is in general not equal to $\beta(a)$ or $\gamma(a, a^\prime)$.

\section*{Standardization does not recover $\beta(a)$}

To see that $\phi(a,a^\prime)$ is not in general equal to $\beta(a)$, we write the latter as follows: 
\begin{equation}\label{eq:beta}
    \begin{split}
        \beta(a) &= \E[Y^a | S = 0] \\
            &= \int \E[Y^a | X = x, S= 0] f(x|S=0)dx \\
            &= \int \E[Y^a | X = x, S= 1] f(x|S=0)dx \\
            &= \int \E[Y^a | X = x, S= 1, A = a] f(x|S=0)dx \\
            &= \int \E[Y | X = x, S= 1, A = a] f(x|S=0)dx.
    \end{split}
\end{equation}
This result, versions of which have appeared in a number of previous contributions (e.g., \cite{pearl2011, westreich2017, dahabreh2020transportingStatMed}), establishes that $\beta(a)$ can be written as a function of the observable variables that standardizes the conditional outcome mean under treatment $A = a$ in the trial, $\E[Y | X = x, S= 1, A = a]$, over the distribution of the covariates among individuals in the target population, $f(x|S=0)$, regardless of treatment use. 

By comparing equation \eqref{eq:statistical_parameter_general} and the last row of equation \eqref{eq:beta}, we see that $\phi(a,a^\prime)$ will not in general equal $\beta(a)$ because they standardize to two different distributions -- $f(x|S=0, A = a^\prime)$ and $f(x|S=0)$, respectively -- that are in general not the same when the measured covariates $X$ are associated with treatment $A$ in the target population ($S = 0$). Thus, estimators of $\phi(a,a^\prime)$ produced by standardization procedures after conditioning on treatment $A = a^\prime$ will be biased for $\beta(a)$.

In our example, the bias would occur if the covariate distribution of warfarin users were unrepresentative of the covariate distribution of all individuals in the target population, that is, when $f(X|A=0,S=0) \neq  f(X|S=0)$. As a result, standardization of the trial results to the warfarin users would not produce the same results as standardization of the trial results to all individuals in the target population. The latter would indeed estimate the expectation of the counterfactual outcome in all non-randomized individuals in the target population (i.e., $\beta(a)$) under the causal DAG of Figure 1 (A), provided the identifying assumptions listed above hold. But such analyses would require using covariate information from a representative sample of the entire target population, not just the subset using warfarin.

The bias arises whenever there is an open path between covariates $X$ and treatment $A$ conditioning on $S=0$ in the DAG (Figure 1C), i.e., whenever $X \nindep A|S=0$. In Appendix Figure 2 we give some example causal structures that illustrate that the bias can occur even in the absence of unmeasured confounding of the treatment -- outcome association outside the trial. For example, the bias can occur when (i) covariates $X$ have a direct effect on treatment $A$ (Appendix Figure 2, Panel (A)), (ii) covariates $X$ and treatment $A$ share a unmeasured common cause (Appendix Figure 2, Panel (B)), or (ii) when there is an unmeasured common cause of trial participation $S$ and treatment $A$, such that $X$ will be associated with $A$ in the target population with $S=0$ ( Appendix Figure 2, Panel (C)).  

In some special cases, $\phi(a,a^\prime)$ can equal $\beta(a)$. For example, suppose that starting with the DAG of Figure 1, we (i) remove the $X \rightarrow A$ edge and the $U \rightarrow A$  edges,  or (ii) remove the $X \rightarrow A$ and the $U \rightarrow X$ edges, such that there is no open path between $X$ and $A$ conditional on $S=0$ (Appendix Figure 3). With either of these modifications, the covariate distribution among a specific treatment group in the the target population is representative of the entire target population (e.g., the covariate distribution of warfarin users will be the same as that of all individuals in the target population), such that $f(X|(A=a^\prime, S=0)= f(X|S=0)$; and therefore,  $\phi(a,a^\prime)$ will equal $\beta(a)$. Furthermore, under the sharp causal null for the treatment effect, standardization methods can recover the average treatment effect in the target population (i.e., $\phi(1,a^\prime)-\phi(0,a^\prime) = \beta(1)-\beta(0) = 0$), but will not not recover the counterfactual expectations under different treatments in the target population (i.e., $\phi(a,a^\prime)$ will in general not equal $\beta(a)$, for $a=0,1$).

\section*{Standardization does not recover $\gamma(a,a^\prime)$}

Even though $\phi(a,a^\prime)$ does not in general equal $\beta(a)$ under the causal DAG in Figure 1, we might hope that $\phi(a,a^\prime)$ equals $\gamma(a,a')$, because both of these parameters pertain to the subset of the target population using treatment $A = a^\prime$. 

To see that $\phi(a,a^\prime)$ is not in general equal to $\gamma(a,a')$, we begin by rewriting the latter as follows: 
\begin{equation}\label{eq:gamma}
    \begin{split}
     \gamma(a,a') &= \E[Y^{a} | S = 0, A = a'] \\
     &= \int \E[Y^{a} |X = x, S = 0, A = a'] f(x | S = 0, A = a^\prime)dx.
    \end{split}
\end{equation}
By comparing equation \eqref{eq:statistical_parameter_general} and the second row in display \eqref{eq:gamma}, we see that $\phi(a,a^\prime)$ will not in general equal $\gamma(a,a^\prime)$ because $ \E[Y^{a} |X = x, S = 1, A = a']$ and $\E[Y^{a} |X = x, S = 0, A = a']$ are not in general equal when the counterfactual outcomes $Y^a$ are not independent of trial participation $S$ given covariates and treatment, that is, when $Y^{a} \not\independent S | (X,A)$. Thus, estimators of $\phi(a,a^\prime)$ produced by standardization procedures after conditioning on treatment $A = a^\prime$ are biased for $\gamma(a,a')$. 

The causal DAG of Figure 2 shows that this bias is due to conditioning on a collider (collider stratification). In the Figure, we represent selection on treatment by placing a box around the corresponding node. Treatment $A$ is a collider on the $S \rightarrow \boxed{A} \leftarrow U \rightarrow Y$ path and conditioning on it opens (unblocks) the path between trial participation $S$ and the outcome $Y$; thus, we do not expect the independence $Y^a \independent S | (X,A)$ to hold. Consequently, we also do not expect $ \E[Y^{a} |X = x, S = 1, A = a']$ to equal $\E[Y^{a} |X = x, S = 0, A = a']$ in general and we also do not expect $\phi(a,a^\prime)$ to equal $\gamma(a,a')$, even though these two parameters involve marginalization over the same covariate distribution. 

To gain some intuition about the above results in the context of our running example, suppose $U$ is an indicator of participants' stroke risk (high vs. low) under warfarin treatment, unexplained by the measured covariates. Suppose also that outside of the trial ($S=0$), individuals with high stroke risk under warfarin are more likely to use dabigatran and also more likely to develop stroke, represented by the fork $A$ $\leftarrow$ $U$ $\rightarrow$ $Y$. In other words, outside of the trial, individuals who use warfarin are more likely to be low-risk individuals compared with those who use dabigatran. In contrast, among trial participants ($S=1$), individuals who are assigned to warfarin are exchangeable with those who are assigned to dabigatran (by randomization). Furthermore, suppose that treatments are assigned at 1:1 ratio in the trial ($S=1$) but that, within levels of the covariates $X$, warfarin treatment is much more common than dabigatran treatment in the target population ($S=0$), because dissemination of dabigatran into practice soon after the completion of the trial is relatively limited. The difference in utilization rates and treatment assignment mechanisms, within levels of $X$, is represented by the $S \rightarrow A$ edge. Even under the sharp null hypothesis for the treatment effect, within levels of the measured covariates $X$, non-trial participants who use warfarin will on average be at lower risk of stroke under warfarin treatment (as determined by $U$) compared with trial participants assigned to warfarin. Therefore, within levels of the measured covariates, the (counterfactual) stroke risk under warfarin treatment among trial participants receiving warfarin will be on average lower than the same risk among non-participants receiving warfarin (because of the imbalance between the two groups in terms of the unmeasured risk indicator $U$).

In some special cases, $\phi(a,a^\prime)$ can equal $\gamma(a,a^\prime)$. Two such special cases are of particular interest: starting with the causal DAG in Figure 1, suppose that we (i) remove the $U \rightarrow Y$ edge, or (ii) remove $U \rightarrow A$ edge (see Appendix Figure 4). With either of these modifications, conditioning on $A$ will not induce collider stratification bias. Informally, we can say that $\phi(a,a^\prime)$ will equal $\gamma(a,a^\prime)$ when there is no unmeasured confounding of the treatment -- outcome association in the target population ($S=0$). This is an unlikely situation in practice because generalizability and transportability analyses are typically undertaken precisely when observational studies are unreliable due to the presence of confounding of the treatment -- outcome association in the target population. Furthermore, similar to what we noted about estimands that pertain to the entire target population, under the sharp causal null for the treatment effect, standardization methods can recover the average treatment effect in the subset of the target population using treatment $A = a^\prime$ (i.e., $\phi(1,a^\prime)-\phi(0,a^\prime) = \gamma(1,a^\prime)-\gamma(0,a^\prime) = 0$), but cannot recover the counterfactual expectations under different treatments in the target population (i.e., $\phi(a,a^\prime)$ will in general not equal $\gamma(a,a^\prime)$, for $a=0,1$).

\section*{Weighting-based Approaches Have the Same Limitations}

Many applied generalizability and transportability analyses use weighting-based estimators rather than marginalization-based estimators \cite{dahabreh2019relation}. As noted, for outcomes measured at the end of the study, the marginalization-based estimator after conditioning on treatment in the target population is the sample analog of $\phi(a,a^\prime)$. The problems and biases we described above, however, apply equally to weighting estimators. In eAppendix B, we show that the expressions in the main text of the paper have algebraically equivalent weighting representations; commonly used weighting estimators are sample analogs of the weighting expressions we present in the eAppendix. Thus, we can conclude that the issues we highlighted can affect all methods of estimating causal effects in the target population, including marginalization-based, weighting, and doubly robust estimators, when applied to samples obtained by conditioning on treatment.

\section*{Numerical Illustration}

We now consider a numerical example of a hypothetical cohort of 100,000 individuals simulated using a distribution for $(X,S,A,U,Y)$ compatible with the causal DAG of Figure 1. We provide details about the models used to generate data in eAppendix C. For simplicity, all variables were binary. Table 1 shows the data on the measured variables $(X, S, A, Y)$ from this hypothetical cohort. Data on $U$ are not measured (e.g., would not be available to an analyst); in the simulation, however, we know the joint distribution of the the measured and unmeasured variables and can use that knowledge to determine the parameter value for $\gamma(a, a^\prime)$. The parameters $\beta(a)$ and $\phi(a,a^\prime)$ can be written in terms of the measured variables and do not require information on $U$. We obtained the parameter values for all parameters by writing them in terms of the (known) parameters of the simulation model. 

In our simulation, the parameter values of the counterfactual expectations in the target population were $\beta(a=1) = 38.6\%$ and $\beta(a=0)=32.2\%$, and the risk difference in the target population was 6.4\%. Furthermore, the parameter values of the expectations of the counterfactual outcome in the subset of the target population using treatment $A = a^\prime = 0$ were $\gamma(a=1,a'=0) = 28.2\%$ and $\gamma(a=0,a'=0)=43.2\%$, corresponding to a risk difference of -14.9\%.

Using the data in Table 1 we obtained plug-in estimates of $\phi(a,a^\prime)$ using equation \eqref{eq:statistical_parameter_general} (here, we use ``hats'' to indicate estimates). Readers can verify the following calculations using the data in the table:  $\widehat\phi(a=1,a^\prime=0) = 35.0\%$ and $\widehat\phi(a=0,a^\prime=0) = 35.1\%$, corresponding to a contrast of $\widehat\phi(a=1,a^\prime=0) - \widehat\phi(a=0,a^\prime=0) = -0.1\%$. These values are grossly different from the ``true'' values of the causal estimands given in the previous paragraph, both those pertaining to the target population and those pertaining to the subset with $A =0$. Thus, we can conclude that conditioning on treatment when sampling the target population can potentially induce significant bias in the estimation of counterfactual expectations and treatment effects. Table 2 summarizes the numerical results to facilitate comparisons.

\section*{DISCUSSION}\label{sec:discussion}
In this paper, we describe previously unappreciated identification problems and biases in studies that attempt to extend causal inferences from a trial to a target population when the target population is sampled conditional on treatment. We showed that under a simplified causal model, conditioning on treatment and applying common standardization methods results in the statistical estimand that in general is not equal to either (1) the expectation of the counterfactual outcome in the target population, or (2) the expectation of the counterfactual outcome in the subset of the target population using a specific treatment.

The sources of bias for common analytic approaches differ with respect to each of these causal quantities of interest. When the causal quantify of interest is the expectation of the counterfactual outcome in the target population, that is, $\beta(a)$, bias arises because of an open path between covariates $X$ and treatment $A$ among non-randomized individuals ($S=0$). The open path can be produced by a direct effect of the covariates on treatment; unmeasured common causes of the covariates and treatment; or by conditioning on a common effect (as in Appendix Figure 2, Panel C) \cite{hernan2004structural}. The existence of such a path makes the covariate distribution of the non-randomized individuals who use a specific treatment unrepresentative of all non-randomized individuals in the target population.

In contrast, when the target causal quantity of interest is the expectation of the counterfactual outcome in the subset of the target population using $A = a^\prime$, that is, $\gamma(a,a^\prime)$, bias arises because treatment is a collider on a path between trial participation and the outcome. This form of collider stratification bias makes individuals in the trial non-exchangeable with individuals in the target population who use a specific treatment or a subset of the treatments. The structure of the bias is similar to the bias induced by selection on a subset of the possible treatments in instrumental variable analyses \cite{swanson2015selecting,  ertefaie2016selection}. This is not surprising because, in the causal DAG of Figure 1, trial participation $S$ is an instrument for the effect of treatment $A$ on the outcome $Y$, conditional on covariates $X$.

Our results have implications for studies combining evidence from difference data sources to estimate treatment effects. First, our findings suggest that for most generalizability or transportability analyses, sampling the target population should not depend on treatment. Because individuals whose information is captured in routinely collected data can meet the trial eligibility criteria at multiple times, future work should evaluate strategies for sampling the target population. 

Second, the problems engendered by conditioning on treatment in the target population when combining a randomized trial (or a single arm experimental study) with observational data in order to compare an experimental treatment against a control treatment used in the observational data, regardless of the exact method of comparison. For example, the problems we identify also affect studies using so-called matching adjustment indirect comparison (MAIC) methods \cite{signorovitch2012matching, jackson2021alternative} or any other external comparator method \cite{mishra2022external}, when the data from the target population are selected on the basis of treatment subject to confounding by unmeasured variables.

Third, methods of extending inferences from a trial to a target population are an important component of benchmarking observational analyses against trials \cite{dahabreh2020benchmarking}. Our findings suggest that in benchmarking attempts, restricting the observational data to individuals receiving the treatments evaluated in the trial (when there exist additional treatments in the target population) results in estimates that do not pertain to the entire population underlying the observational data and complicates the interpretation of findings.

Fourth, at a conceptual level, our results highlight the importance of viewing generalizability and transportability analyses as problems where the causal structure needs to be explicitly considered. Such analyses are often thought as straightforward applications of standardization methods to a distribution that can be estimated by treating any sample of non-randomized individuals as a sample from the target population. Our results suggest that naive application of standardization methods to samples selected conditional on treatment can be problematic under plausible causal structures that allow for the presence of confounding in the target population. 

In summary, in studies that attempt to generalize or transport causal inferences from a trial to a target population, conditioning on treatment complicates identification and can lead to bias. We hope that our findings will lead to a critical re-evaluation of the common practice of conditioning on treatment in applied generalizability and transportability analyses.

\section*{ACKNOWLEDGMENTS}



This work was supported in part by Patient-Centered Outcomes Research Institute (PCORI) award ME-1502-27794 (Issa Dahabreh) and American Heart Association (AHA) grant \#834106 (Yu-Han Chiu).  \\
The content of this paper is solely the responsibility of the authors and does not necessarily represent the official views of PCORI, PCORI's Board of Governors, the PCORI Methodology Committee, or the AHA. 


\clearpage
\section*{TABLES}\label{sec:Table}
%
 
\begin{table}[htp]
\caption{Data on information source $S$, covariate $X$, treatment $A$, and outcome $Y$ in a simulated example.}\label{tab_observed_data}
\renewcommand{\arraystretch}{1.5}
\begin{tabular}{|c|c|c|cc|} 
\hline
Information source     & Covariate          & Treatment & \multicolumn{2}{c|}{Outcome}     \\ \hline
$S$                  & $X$                  & $A$         & \multicolumn{1}{c|}{$Y = 1$}     & $Y = 0$  \\ \hline
\multirow{4}{*}{1} & \multirow{2}{*}{1} & 1         & \multicolumn{1}{c|}{2986}  & 3669  \\ \cline{3-5} 
                   &                    & 0         & \multicolumn{1}{c|}{1630}  & 4793  \\ \cline{2-5} 
                   & \multirow{2}{*}{0} & 1         & \multicolumn{1}{c|}{4637} & 17536 \\ \cline{3-5} 
                   &                    & 0         & \multicolumn{1}{c|}{11097}  & 11501 \\ \hline
\multirow{4}{*}{0} & \multirow{2}{*}{1} & 1         & \multicolumn{1}{c|}{10206}  & 11276  \\ \cline{3-5} 
                   &                    & 0         & \multicolumn{1}{c|}{3191} & 6201 \\ \cline{2-5} 
                   & \multirow{2}{*}{0} & 1         & \multicolumn{1}{c|}{1492}  & 3258  \\ \cline{3-5} 
                   &                    & 0         & \multicolumn{1}{c|}{3669}  & 2858  \\ \hline
\end{tabular}
\end{table}

\vspace{1in}

\begin{table}[htp]
\caption{Estimates for different parameters using data from Table \ref{tab_observed_data}.}
\renewcommand{\arraystretch}{1.5}
\begin{tabular}{|l|c|c|c|c|}
\hline
       Statistical parameter        & $a = 1$ & $a = 0$ & Estimated difference & Estimated ratio \\ \hline
$\widehat \phi(a,a^\prime=0)$   & 35.0\%                 & 35.1\%    & -0.1\%            & 1.00       \\ \hline
$\beta(a)$        & 38.6\%                 & 32.2\%           & 6.4\%     & 1.20       \\ \hline
$\gamma(a,a^\prime=0)$ & 28.2\%                 & 43.2\%           & -14.9\%         & 0.65       \\ \hline
\end{tabular}
\caption*{Values for $\gamma(a,a^\prime=0)$ are obtained using information on $U$ which would not be available to analysts in practical applications, but are available to us in the simulation.}
\end{table}

\clearpage
\section*{FIGURES}\label{sec:Figures}

\begin{figure}[htp]
 \caption{Directed acyclic graphs (DAGs) for a simplified causal model for extending inferences from a trial to a target population. Figure 1A represents data from both randomized ($S=1$) and non-randomized ($S=0$) individuals. Figures 1B and 1C are conditional on trial participation (indicated by placing a box around the $S$ node).}
    \vspace{0.5in}
    \centering
 \includegraphics[width=20cm]{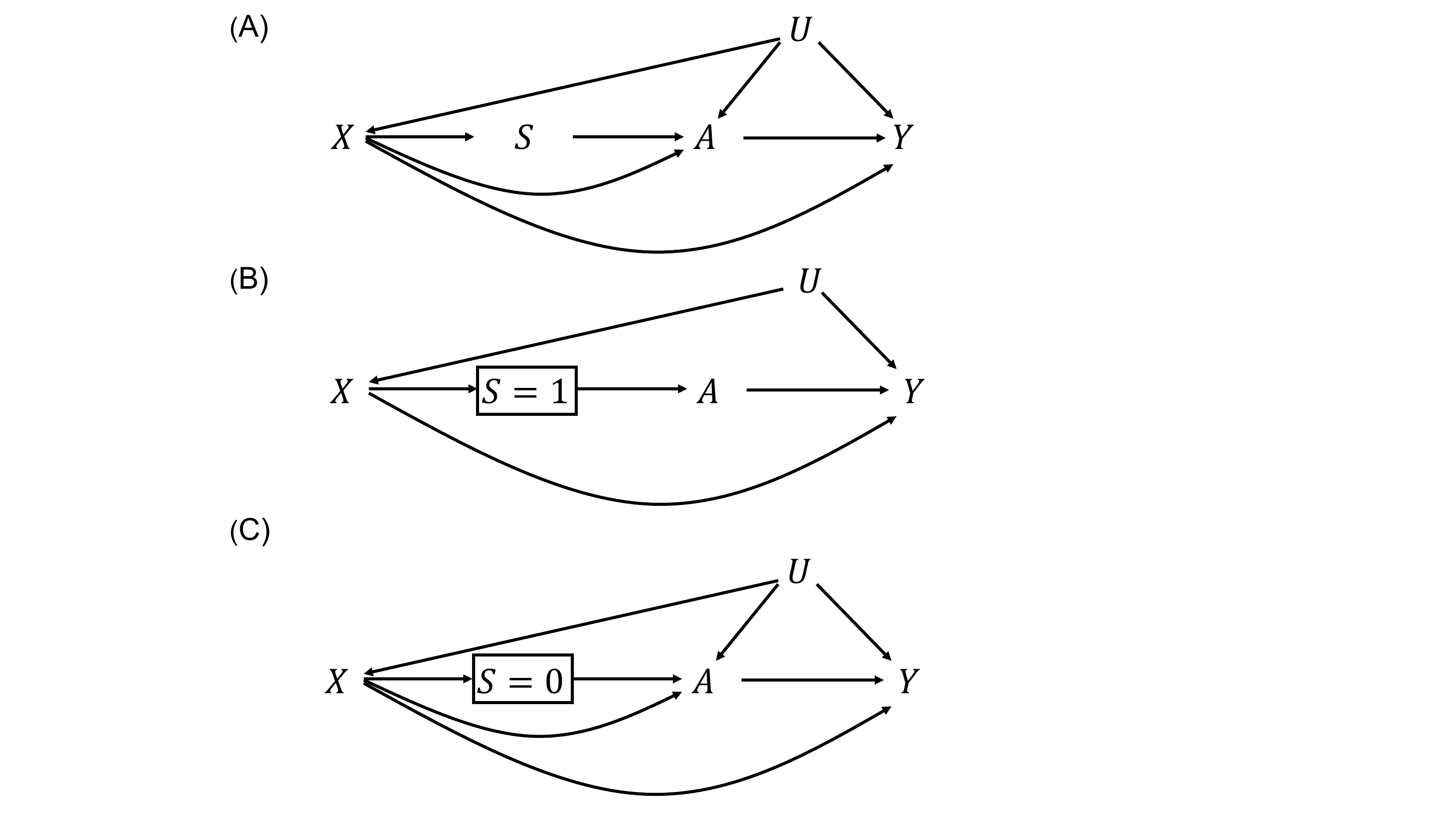}
      \label{fig:dag_one}
\end{figure}

\begin{figure}[htp]
 \caption{Directed acyclic graphs (DAG) obtained by conditioning on treatment in the model depicted in Figure 1A.}
    \vspace{0.5in}
    \centering
 \includegraphics[width=20cm]{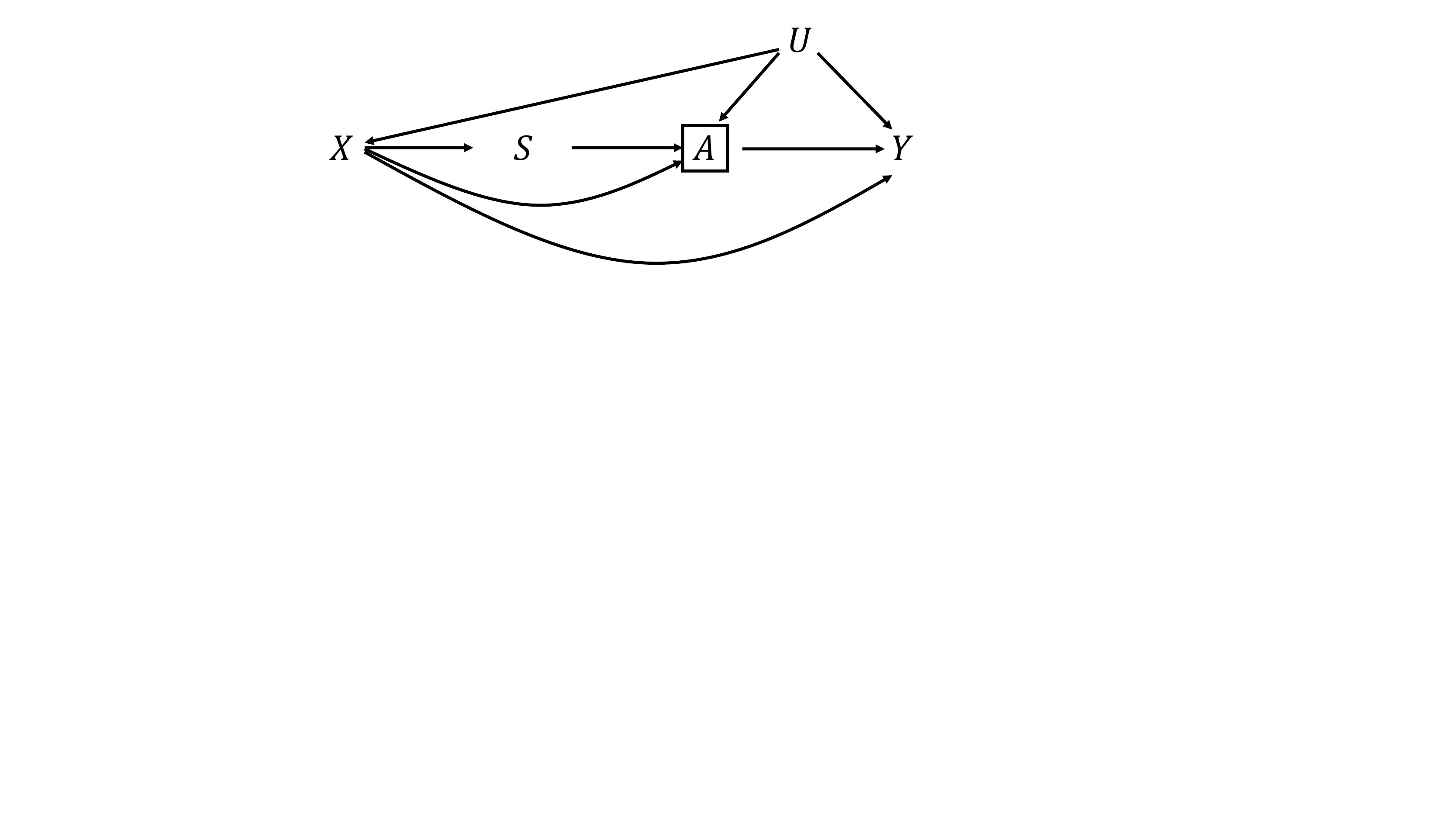}
      \label{fig:dag_two}
\end{figure}

\clearpage
\bibliographystyle{unsrt}
\bibliography{bibliography}{}


\ddmmyyyydate 
\newtimeformat{24h60m60s}{\twodigit{\THEHOUR}.\twodigit{\THEMINUTE}.32}
\settimeformat{24h60m60s}
\begin{center}
\vspace{\fill}\ \newline
\textcolor{black}{{\tiny $ $selection\_on\_treatment, $ $ }
{\tiny $ $Date: \today~~ \currenttime $ $ }
{\tiny $ $Revision: \paperversionmajor.\paperversionminor $ $ }}
\end{center}

\end{document}